\documentclass[aps]{revtex4}

 \newcommand{\be}{\begin{equation}}
\newcommand{\ee}{\end{equation}} \newcommand{\rf}[1]{(\ref{eq:#1})}

\usepackage{graphicx}

\begin{document}

\input{psfig.tex}

\title{Velocity Curves for Stars in Disk Galaxies:\\A case for  \\ Nearly Newtonian Dynamics } \author{ M. D. Maia$^{[\ast]}$ }
\author{A. J. S. Capistrano$^{[\dag]}$ } \author{D.
Mulller$^{[\ddag]} $ }
\affiliation{ Universidade de Bras\'{\i}lia,
Instituto de F\'{\i}sica, Bras\'{\i}lia,
DF.70919-970,\\{[$\ast$]maia@unb.br},
{[$\dag$]yishaicapistrano@yahoo.com.br},{[$\ddag$]muller@fis.unb.br}}
\begin{abstract}
The  dark matter  constraint   imposed by the recent  WMAP experiment on  gravitational  theories  is    analyzed.   Using  the non-linearity of  the  vacuum  Einstein's  equations, it is  shown that  when  the slow  motion condition is applied to  the geodesic  equations,   the resulting   nearly Newtonian    gravitational field describes  nearly flat velocity  curves for rotating   stars in the  vicinity      of thin disk galaxies.
  \end{abstract}

 \maketitle

\section{Introduction}

The discrepancy between the observed rotation velocity curves
for stars in a spiral galaxy
 and the theoretical prediction from Newtonian mechanics is a
 long standing problem in modern astrophysics.
  The velocity measurements are based on
the Tully-Fischer relation between the mass of the galaxy and the
width of the 21-cm line of hydrogen emissions, suggesting that a
larger galaxy mass would increase the rotation rate.
  Since these velocities are much smaller
than the speed of light $c$, in principle they should be
described by Newton's gravitational theory. However, as shown in
the example of Figure 1 for the galaxy NGC3198, the theoretical
Newtonian curve agrees with the experimental one only at the
galaxy's nucleus \cite{Albada}. For larger distances, the observed curve becomes almost horizontal, separating from the theoretical
Newtonian curve which drops rapidly with $r$. Such pattern is
observed in most spiral galaxies and galaxy clusters
\cite{Yoshiaki,Moffat}.

\begin{figure}[!h]
\includegraphics[width=3.6in]{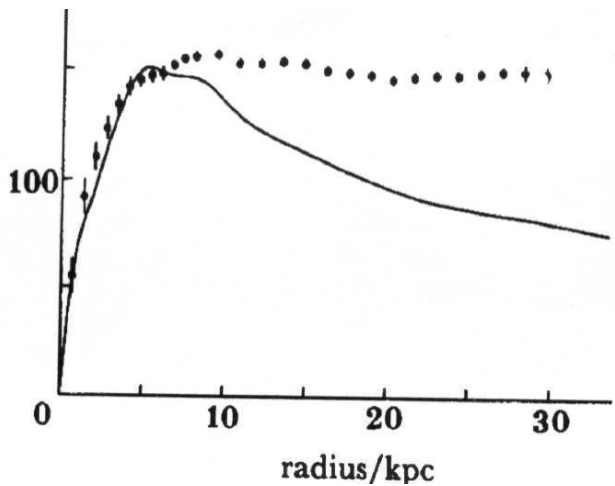}
\caption{ The Observed Rotation velocity curve (error bars)
 compared with the predicted Newtonian curve of the NGC3198.  }
\end{figure}

The most common explanation   for this  problem was originally
proposed by Zwicky in 1933 \cite{Zwicky}.
 Accordingly, a certain quantity of dark  matter,  invisible with respect to the electromagnetic radiation
spectrum, should be added to each galaxy. Such matter can in principle be composed of ordinary baryonic matter, like planets
distributed in a spherical halo orbiting the galaxy itself, far away from the stars \cite{Kolb}. These have   been  observed with the help of the gravitational microlensing effects, but only in very small amounts, far beyond the required quantity to correct the velocity curves.  In the  cosmological scale,  dark matter   seems  to be  consistent   with  the standard  FRW  model \cite{Khalilova},  but only   recently   the  cosmic microwave radiation  data  analysis from the  WMAP  experiment indicated that   most of the estimated 22\% dark matter  content   of the universe must be   of
non-baryonic nature.  More  specifically, the analysis of the power  spectrum indicates   that     a theory   of gravity   based essentially  on  the properties of   baryonic  matter  would produce a lower  third   peak \cite{Spergel}. Therefore, either   some  exotic particles  must be  considered \cite{Primack},  or   else  an  adequate  gravitational  theory   should be devised.

In principle that constraint does not exclude non-linear theories like general  relativity. However, general relativity has such  strong  commitments  with  its   Newtonian  limit, that it makes it  difficult to explain the rotation velocity curves. The usual  argument goes as  follows:  The velocity curves for  stars in a spiral galaxy derived from Newton's theory agree with the observed curves only at the galaxy's core  (FIG. 1.), precisely where the space-time curvature produced by  Einstein's gravity would be more pronounced. Beyond that region, the gravitational field becomes sufficiently weak to be taken over by its  Newtonian limit.
Over   the  time, this  has  motivated  research on  many  alternative  gravitational theories, which we  separate into   two main categories:

\vspace{2mm}

\textbf{(i)} \textbf{Modifications of  Newtonian Gravity}\\
According to this proposal, Newton's gravitational theory should be modified so as to correctly describe the velocity curves.
The first  thought is  of  course the  post-Newtonian approximations of general relativity,  regarded  as  corrections to  Newtonian theory. However,  a simple exercise shows that  in a second order   parametric post-Newtonian approximation,  the  correction term   in the velocity curves decay  with  $1/r^3$,  not improving  substantially the  velocity curves \cite{Fernando}.  On the other  hand,   post-Newtonian  cosmology (see  e.g. \cite{Szekeres1})  does not meet the  WMAP power  spectrum  constraint.

Other   modifications of the Newtonian theory  have  been  considered \cite{Mannheim,Moffat,Milgrom}. Among these, MOND has received a substantial attention and it has been  backed by a theory in
which Poisson's equation for the Newtonian gravitational field is replaced by an equation like \cite{Bekenstein} \[ <\nabla, \mu(\frac{|\nabla\varphi|}{a_{0}})\nabla\varphi> =4\pi G\rho \] where $\mu (x) $ is a function to be adjusted to the specific type of galaxy and $a_{0}$ is a constant acceleration. For example, in a
spherically symmetric distribution of matter, it is suggested that $\mu (x)=x/\sqrt{x^{2}-1}$, producing the following correction for Newtonian potential
\[
\varphi =\sqrt{a_0 G M} \ln r
\]
This theory has shown good agreement with most known spiral
galaxies, but  there are reports suggesting that it may be constrained by  galaxy clusters
\cite{Pointecouteau,Shirata}.  Finally, in  spite of being essentially a  local  gravitational theory, its global   effect  on the  composition of the total energy of the universe   does not  meet  the power  spectrum  constraint  \cite{Spergel}.
\vspace{2mm}

 \textbf{(ii)} \textbf{Modifications of  General Relativity}\\
 There are just too many ideas on how to modify general relativity  to correct the velocity curves based on  a variety  of  suppositions. Here we just list some  of these: (1) Add a scalar field to Einstein's equation, in such a way that the scalar-tensor theory corrects the Newtonian limit
\cite{Fay}; (2) Modify the concept of time in general relativity, so that the Newtonian limit of the theory differs from the original
Newton's' theory \cite{Carmeli,Hartnett}; (3) Add a cosmological
constant with the appropriate sign (depending on which side of the equation it is placed) \cite{Whitehouse}; (4) Include higher order curvature terms in the gravitational variational principle as a
means to increase the local gravitational pull on galaxies
\cite{Cappozielo}; (5) Several brane-world models and variants have been considered,  in the hope that   the additional degree of freedom would explain the rotation curves.
\cite{Mak,Vollick,Takeshi,Nobuchika,Ichiki,Lidsey,Cembranos}.

\vspace{4mm}

The  purpose of this paper is  to  show   that   when the  slow  motion  condition $v<<c$  is   applied to the   geodesic  equations  only,  then the  self interacting   vacuum gravitational  field produced   by  a  disk  galaxy,  contributes to   a  nearly   Newtonian   motion of  a star  in the galactic  plane,  with nearly  flat  velocity curves.

This  is  justified  first  by   the  fact that  the  geodesic  equations  are  derived  from   Einstein's  equations,  but in the Newtonian  limit the  equations of  motion   corresponds to  a  separate postulate \cite{Infeld}. Therefore,  when   the gravitational  field   of  a galaxy  acting upon a free  falling star  is   sufficiently weak,  then   the  slow  motion  condition   $v<<c$  applies to Einstein's  equations and the   only remaining   option  is the Newton's  law of  motion.  On the other hand,  admitting  that  the  free   falling star gets   in  a   region  where   the  gravitational  field  is  beyond the  Newtonian limit, then  the  condition  $v<<c$  still applies  to   the  geodesic  equations, but  not necessarily on  Einstein's  equations.   In fact, the  geodesic  equations depend only linearly   in  the connection,  while   Einstein's equations  depend  quadratically  in the same  connection.  Therefore,   the effect  of the  condition  $v<<c$  in the connection,  becomes  less restrictive  on the geodesic  equation than   in  Einstein's  equations. The result  is   that in that  region   a  nearly   Newtonian gravity   prevails.

\section{Nearly Newtonian Gravity}

Consider a slowly free falling star, $v<<c $, in a region of the space-time, where the pull of the gravitational field on the particle is initially weak:
 \begin{equation}
 g_{\mu\nu}=\eta_{\mu\nu} +\delta h_{\mu\nu},\;\;\;
  \delta h^2_{\mu\nu}<<\delta
 h_{\mu\nu},\label{eq:weak}
 \end{equation}
\underline{where $\delta h_{\mu\nu}$ is not parameterized by $v/c$}.
Under these conditions Newtonian coordinates, can be applied, so
that the three spatial components of the geodesic equation become
(Here  we  follow  essentially the  derivation in \cite{Wheeler})
   \be
 \frac{d^2
x^{i}}{dt^2} =
 -\Gamma^{i}_{ij} \frac{d x^{i}}{dt}
 \frac{dx^{j}}{d t} -2\Gamma^{i}_{j 4} \frac{d x^{j}}{d t}=
 -\Gamma_{44}^{i} = -\frac{1}{2}\delta_{44,i}\label{eq:geodesic}
 \ee
 where  $t$  denotes  the  Newtonian  time. Therefore,  we may
apply a Newton's-like
 equations of motion  for  a scalar field $\phi$ defined by
  \be
 \frac{d^2 x^{i}}{dt^2} =
   -\frac{\partial\phi}{\partial x^{i}}
 \label{eq:newton}
\ee
Notice that  $\phi$ is not necessarily the Newtonian potential
because  the  $v<<c$ condition  was not    applied to  Einstein's  equations. Comparing the above expression with \rf{geodesic}, we obtain
 \be
\frac{\partial\phi}{\partial x^{i}} =
-\frac{1}{2}\frac{\partial\delta h_{44}}{\partial x_i}
\label{eq:smallphi}
\ee

 As the particle continues  its free fall, while maintaining the slow motion,  the gravitational field continuously
 builds up by small increments as
 $$ g_{\mu\nu} \approx \eta_{\mu\nu} +\delta h_{\mu\nu}+ (\delta
 h_{\mu\nu})^2 +\cdots$$
Actually, there is no reason to stop this process, so that
\rf{smallphi} can be integrated for all perturbations of the
Minkowski metric, up to a finite $h_{\mu\nu} $, leading to
 \be
 \phi= -\frac{1}{2}\int_{0}^{h_{44}} d(\delta h_{44})=
-\frac{1}{2}(1+g_{44}) \label{eq:NND}
\ee
 This nearly Newtonian gravitational  potential is characterized
 by an exact solution of Einstein's equations,  with the non-linear effects  present in the component $g_{44}$   \cite{Wheeler}.

\vspace{3mm}
In order to understand the implications of \rf{NND}  to the dark matter problem, suppose  that we have  the Schwarzschild's solution of Einstein's equations written in the
usual spherical coordinates,  so that  $g_{44} =-(1 -2M/r)$.
 In this case,  \rf{NND} coincides with Newton's gravitational potential $\phi =-M/r$ for a spherically symmetric gravitational source with mass $M$.
If, this particular potential is  applied  to describe the motion of a star in a spiral galaxy  corresponding to  a   spherically symmetric  "visible mass" $M$, it does not describe correctly the rotation curves  outside the  galaxy  nucleus,  regardless of how strong the Schwarzschild field may be.
On the other hand,  if the star is  close to the galaxy  nucleus, then  it will  feel the pull of a spherically symmetric  gravitational field which  coincides  with the above Newtonian potential $-M/r$. This   coincidence  explains  why  the  two  curves in  Fig.1  agree  at the galaxy's nucleus.
 For  any other   solution of  Einstein's  equations which is not diffeomorphic  to   the  Schwarzschild's   solution,  \rf{NND}  will  produce a  different near Newtonian dynamics.  Our  understanding is that the  nearly Newtonian potential  \rf{NND} carries a symmetry dependent    non-linear effects contained in Einstein's  equations  through the  component  $g_{44}$, as  it will be exemplified  in the  next section.

It is  relevant  to distinguish the  present application of    \rf{NND}  from a  solution of   the same   problem  using full general  relativity as  in \cite{Garfinkel,Cooperstock}. Here, besides  having lost   general covariance as  a  consequence of  the slow  motion condition,  only one  component of  metric has  a direct contribution to the motion. In the following section we show that the velocity  curves   derived from  \rf{NND}   for a  vacuum gravitational field are   compatible with the observed  curves, using an exact solution of the vacuum Einstein's equations corresponding to a disk galaxy.

\section{Velocity Curves near a Disk Galaxy}

As a simple model for a disk galaxy we may consider a cylinder such
that its height $h_0$ is much smaller than its radius $r_0$. The
line element produced by such object can be derived from the Weyl
cylindrically symmetric metric, expressed in cylindrical coordinates
$(r,z,\theta)$ as \cite{Weyl}
\begin{equation} dS^{2}
=e^{2(\lambda-\sigma)}dr^{2} + r^{2}e^{-2\sigma}d\varphi^{2}
+e^{2(\lambda-\sigma)}dz^{2} - e^{2\sigma}dt^{2} \label{eq:weyl}
\end{equation} where $\lambda =\lambda{(r,z)}$ and $\sigma
=\sigma{(r,z)}$. The exterior gravitational field outside the
cylinder, is given by vacuum Einstein's equations:
 \begin{eqnarray}
 && - \lambda_{,r} + r\sigma_{,r}^{2} -r
\sigma_{,z}^{2} =0
\label{eq:first}\\
 &&-\sigma_{,r}-r\sigma_{,rr}- r\sigma_{,zz}=0 \label{eq:second}\\
&& \lambda_{,rr} +\lambda_{,zz} +\sigma_{,r}^2 + \sigma_{,z}^2
=0\label{eq:third}\\ && 2r\sigma_{,r}\sigma_{,z}
=\lambda_{,z}\label{eq:fourth}
\end{eqnarray}

To the above metric we apply the thin disk condition \be z\in
[-h_0/2, h_0/2],\,\,\; \mbox{for} \,\,\; r\in [0,r_0],\,\, h_0<<r_0
 \label{eq:thin}
\ee
 In this case
we may expand the functions $\sigma(r,z)$ and $\lambda(r,z)$ around
$z=0$, obtaining
  \begin{eqnarray*}
&&\phantom{x}\hspace{-5mm}\sigma(r,z)
=\sigma(r,0)+ z a(r)...\\
 &&\phantom{x}\hspace{-5mm}\lambda(r,z)
=\lambda(r,0) + z b(r)...
 \end{eqnarray*}
 where we have denoted
 \be
 a(r)= \left.
  \frac{\partial
\sigma(r,z)}{\partial z}\right\rfloor_{z=0} \;\; \mbox{and}\;\;\,
b(r)=\left.\frac{\partial \lambda(r,z)}{\partial
z}\right\rfloor_{z=0} \label{eq:ab}
 \ee
 The thin disk condition \rf{thin} implies that
  the above expansion can be truncated to the linear terms
   in $z$. Therefore, replacing
$ \sigma_{,zz}=0$ and $\lambda_{,zz}=0$ in \rf{second} and
\rf{fourth}, they become simple ordinary differential equations on
$\sigma$, with general solution $$\sigma(r,z)=\frac{K_0}{2}\ln r +
c_2 (z)$$ where we have denoted
 \be
  K_0 =\frac{b(r)}{a(r)} \label{eq:K}
 \ee
 and where $c_2(z)$ is an r-integration constant.
  Derivation of $\sigma$
 with respect to $z$ gives $c_2(z) =a(r)z + c_0$, but since
  $c_2$
 does not depend on $r$, it follows that
  $a(r)= a_0=$ constant. By similar arguments we find that
  $b (r) =b_0 =$constant, so that $K_0$ is also a constant.
Replacing these results in \rf{first} and \rf{third}, together with
$\lambda_{,zz}=0$, we also obtain ordinary equations for
$\lambda(r,z)$. Therefore, the solution of Einstein's equations for
the thin Weyl disk is
 \begin{eqnarray}
&&\sigma(r,z)=\frac{K_0}{2}\ln r + a_0 z + c_0 \label{eq:sigma}\\
 &&\lambda(r,z) =
\frac{K_0^2 }{2}\ln r - a_0 \frac{r^{2}}{2}+b_0 z + d_0
\label{eq:lambda}
 \end{eqnarray}
 where $d_0$ is another integration constant.

 From \rf{sigma} we
obtain $g_{44}= -e^{2\sigma}=-e^{2\frac{K_0}{2}\ln r} e^{2a_0 z}
e^{2c_0}$. Therefore, for an object moving in the disk plane $z=0$,
we obtain
 \be
 \phi (r)= -\frac{1}{2}(1+ g_{44})\rfloor_{z=0} = -\frac{1}{2}(1
-e^{2c_0} r^{K_0} ) \label{eq:NNDdisk}
 \ee

 In analogy with the
Schwarzschild solution, the integration constant $e^{2c_0}$ may be interpreted as a mass, with the difference that here we cannot
compare it with the same Newtonian mass. However,  we may   assume that  this constant is proportional  to the   visible mass $M_v$ of  a  disk-like galaxy (in  units  G=c=1):  $ e^{2c_0}=2\beta_0 M_v$,  where  $\beta_0$ is a mass  scale factor.  It is even possible  to  interpret the  factor $\beta_0$  as  something  to  do  with the observed  baryonic dark matter,  but  then  we  would require  a  correlation between the  visible and  dark matter  in each galaxy \cite{Paolo}.

The tangent velocity of a star as a function of the distance to the center of the galaxy can now be derived from the Newtonian-like equations of motion \rf{newton}, using the potential \rf{NND}. Taking the force acting upon a star of unit mass with tangent velocity $v=\omega r$, $\omega=$ constant to be $\vec{F}=\frac{v^2}{r}\hat{r}$ and comparing with the force generated by
\rf{NND} $\vec{F}= -\frac{\partial \phi}{\partial r}\hat{r} $, we obtain, $v^2 =r\frac{\partial\phi}{\partial r}$ so that  for  the  considered  disk we obtain ( in units  G=c=1)
 \be
v(r)=\sqrt{|\beta_0 M_v K_0 r^{K_0}|} \label{eq:velocity}
 \ee
The values of $K_0$ given by \rf{K} are determined by the
coefficients of $z$ in the thin Weyl disk metric. Interestingly,
$K_0$ is present even in the plane $z=0$, a subtle consequence of the non-linearity of the vacuum Einstein's equations.

In the thin disk case, the Newtonian velocity can be recovered for $K_0 =-1$ and $\beta_0= 1$. Since this particular value does not contribute to the rotation curves outside
the galaxy core, the value $K_0=-1$ must be ruled out for disk
shaped galaxies. On the other hand, when $K_0 = +1$, the velocity expression \rf{velocity} does not correspond to any observed
rotation curve. We conclude that $|K_0 |$ must be smaller  than  one.

Figure 2 shows the velocities calculated with \rf{velocity} for a few known examples. Since the stars are supposed to be at the rim of each disk galaxy with radius $r_0$, the origins of each curve were shifted, replacing $r$ by $r-r_0$, so that the shown curves start at the estimated $r_0$ for each galaxy. The values of $K_0$ were
determined by comparing the observed average velocity $<v_0>$ with the calculated velocity for each galaxy. In the given examples all values of $K_0$ are positive but different, so that the curves have slightly different slopes. The values of $\beta_0$ do not affect the shapes of the curves and were estimated for each galaxy from the know top speed in each case.

\begin{figure}[h]
\includegraphics[width=9cm]{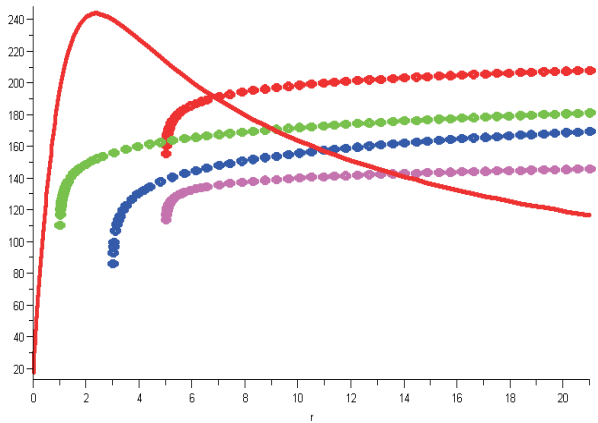}\\
\caption{Velocity curves from \rf{velocity}, for some
examples. Distances are in Kpcs and
 velocities in Km/sec:\\
  (a) The continuous red line
 represents a simulation of the Newtonian
  curve for the Sun in the Milky Way.\\
 (b) Dotted red line is the Milky Way,
 for $\beta_0=1$, $K_0 =0.08$, $M_v = 1\times 10^{11}\times M_\odot$
  and $r_0 =5$ (at the Sun).\\
  (c) Magenta is NGC3198 for $\beta_0=1$, $K_0=0.068$,
 $M_v =6\times10^{11}\times M_\odot$ and $r_0 = 5$.\\
 (d) Green is NGC3949
 for $\beta_0=15.8$, $K_0 =0.13$,
 $M_v =2.5\times 10^{9}\times M_\odot$ and $r_0=1$.\\
 (e) Blue is NGC3877 for $\beta_0=20$,
  $K_0=0.18$, $M=1.1\times 10^{9} \times M_\odot$ and $r_0= 3$}
 \label{multiplot}
 \end{figure}
 For comparison purposes we include below the observed plots
 (error bars)
 for NGC3877 and NGC3949 \cite{Moffat}:
\begin{figure}[!h]
\includegraphics[width=10cm]{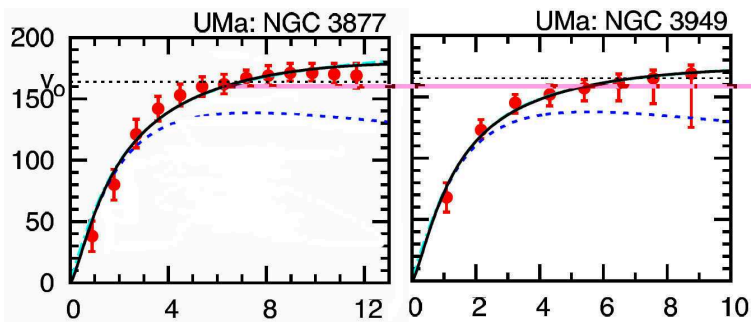}
\caption{Observed rotation curves for NGC3877 and NGC3949}
\end{figure}
\newpage
\begin{center}
\textbf{ Summary}
\end{center}
The slow motion of objects in general relativity is described by
the nearly Newtonian potential, obtained by imposing $v<<c$ in
the geodesic equations only, while leaving Einstein's equations
and the geodesic  deviation equations   intact. The result is the
nearly Newtonian gravity, something   in  between general relativity and Newtonian theory,    characterized  essentially by  $g_{44}$.  The
existence of  such  potential follows from   fact that   in
general relativity the equations of motion are  a  consequence of
the   non-linearity of Einstein's equations, making a  contrast
with Newtonian gravity, where the equation of motion is
postulated separately  from the field  equations.

In particular,  when  the nearly Newtonian potential  is  derived from  a  vacuum solution of Einstein's  equations, the  slow   motion of  a test particle  or  a falling star is   affected by  the  self interaction of the  gravitational field,    so that in principle there are no baryons involved.

The loss  of   general covariance  imposed by   $v<<c$ means  that  the symmetry  of the  gravitational  field solution of the  vacuum Einstein's equations play a significant role in the velocity curves  derived from  \rf{NND}, which is  interpreted  as a  consequence of the non-linearity of Einstein's  equations. In this respect, it should be noted that   the Weyl cylindrical  solution can be transformed to the  Schwarzschild's solution by a diffeomorphism \cite{Rosen}. However,  we  cannot  apply such transformation here because the  diffeomorphism invariance  has been lost. On the other hand, the  solutions of  Einstein's  equations  with a  symmetry that resembles the gravitational field of  a  galaxy  will describe  velocity  curves  which are  closer  to the observed ones.  This was exemplified   by taking the Weyl solution with the format of  a thin disk,  as a model for a disk galaxy.  In this case the     velocity curves  are  remarkably  close  to the experimental curves.

  Clearly, a thin Weyl disk is a very poor mathematical
model for a spiral galaxy. A more realistic  model would be given by a static oblate spheroid, which
can also be derived by a coordinate transformation of the Weyl
metric \cite{Zipoy}. Work on this is still in progress.

\end{document}